\documentclass{emulateapj}
\usepackage{natbib}

\shorttitle{RV Confirmation of a LTT-Detected Binary}
\shortauthors{Barlow, Dunlap, Clemens}

\begin{document}

\title{Radial Velocity Confirmation of a Binary Detected from Pulse Timings\footnote{Based on observations at the SOAR Telescope, a collaboration between CPNq-Brazil, NOAO, UNC, and MSU.}}

\author{B.N. Barlow, B.H. Dunlap, \& J.C. Clemens}
\affil{Department of Physics and Astronomy, University of North 
Carolina, Chapel Hill, NC 27599-3255}
\email{bbarlow@physics.unc.edu}

\slugcomment{Accepted for publication in the Astrophysical Journal Letters}

\begin{abstract}
A periodic variation in the pulse timings of the pulsating hot subdwarf B star CS 1246 was recently discovered via the O-C diagram and suggests the presence of a binary companion with an orbital period of two weeks.  Fits to this phase variation, when interpreted as orbital reflex motion, imply CS 1246 orbits a barycenter 11 light-seconds away with a velocity of 16.6 km s$^{-1}$.  Using the Goodman spectrograph on the SOAR telescope, we decided to confirm this hypothesis by obtaining radial velocity measurements of the system over several months.  Our spectra reveal a velocity variation with amplitude, period, and phase in accordance with the O-C diagram predictions.  This corroboration demonstrates that the rapid pulsations of hot subdwarf B stars can be adequate clocks for the discovery of binary companions via the pulse timing method.
\end{abstract}

\section{Detecting Binaries with Pulse Timings}
\label{intro}

Binary star systems are significant in astronomy because their orbits often permit the calculations of important astrophysical parameters including mass, radius, and density.  Binarity plays an apparently crucial role in the story of the hot subdwarf B (sdB) stars.  Modeled as extended horizontal branch stars in globular clusters, sdBs are presumed to have He-burning cores surrounded by a thin layer of Hydrogen.  Models show they are the progeny of red giant branch stars that somehow lost the majority of their outer Hydrogen layer before reaching the Helium flash.  Many theories attempt to explain the expulsion of this mass, and most of them invoke the presence of a binary companion, either stellar  \citep{han02,han03} or substellar \citep{sok98}.  Observations support this picture; several studies have revealed a large binary fraction amongst the sdBs \citep{max01, cop11}.  Constraining the orbital parameters of these systems may shed light on the evolutionary histories of hot subdwarf stars, and many studies are being carried out with this goal in mind \citep{sch10,muchfuss}.  For further details on hot subdwarfs, we refer the reader to \citet{heb09}.

The majority of hot subwarf orbital parameters calculated to date come from spectroscopic radial velocity (RV) measurements (\citealt{muchfuss}, Table A.1).  In most sdB binary systems, the companion contributes a negligible amount of the system light, so measurements are confined to reflex motion of the sdB.  Although the RV method is quite effective at uncovering reflex motions, the technique is heavily biased towards systems with small separation distances and high masses.  

For binary systems with wider orbits, the light-travel time (LTT) technique\footnote{also known as the pulse timing method when pulsations are used as the clock} provides a viable alternative to the RV method.  As each member in the system orbits the barycenter, the line-of-sight distance from it to the Earth oscillates; thus, light from the object sometimes arrives later than usual and other times earlier.  If the binary contains a pulsating star, this changing light-travel time manifests itself as a phase oscillation in the observed minus calculated (O-C) diagram constructed from the pulse timings.   \citet{sil07} used this technique and found evidence of a 3.2-M$_J$ planet orbiting the sdB star V391 Peg, which demonstrated that some planets can survive the red giant expansion of their host stars.  Several reviews have been published describing the procedures, shortcomings, and successes of this method (e.g., \citealt{ste05}).  

Few systems exist with orbital parameters easily detectable by both the RV and pulse timing methods.  In fact, out of all types of pulsating stars, only two have been observed with RV measurements in accordance with the results of the LTT technique:  AW Persei, a classical Cepheid \citep{vin93},  and SZ Lyn, a $\delta$ Scuti \citep{sze83,bar84,mof88}.  No such corroboration has been demonstrated yet using pulsations with periods on the order of minutes, as exist in the pulsating white dwarf \citep{mon09} and hot subdwarf stars \citep{ost09}.  Nonetheless, much time and effort is being spent accumulating phase measurements of these types of pulsators in hopes of discovering binary members down to planetary-size (e.g., \citealt{sch10, mul08}). 

Recently, we discovered fortnightly phase oscillations in the pulse timings of the sdBV$_{r}$ star CS 1246 via the O-C diagram \citep{bar11}.  An orbiting body with minimum mass of 0.13 M$_{\sun}$ provided the simplest explanation for the phase wobble, but since no secondary light could be detected from the system, this hypothesis remained unconfirmed.  Here, we present corroborating radial velocity evidence for a binary companion around CS 1246.   Our result confirms that the rapid pulsations of hot subdwarf B stars can be sufficient clocks for the detection of companions using the O-C diagram.
 
\begin{figure*}
\epsscale{0.9}
\plotone{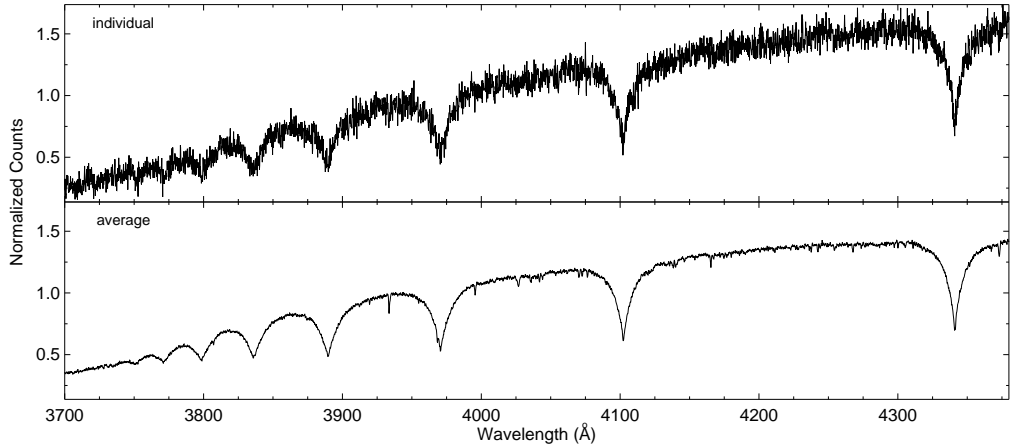}
\caption{SOAR/Goodman spectra of CS 1246.  \textbf{\textit{Top panel:}} Example spectrum used to determine the RV of CS 1246 on a single night.  The spectrum shown, which is a combination of five 371.7-s exposures, has a resolution of 0.52 \AA .  \textbf{\textit{Bottom panel:}}  Average of all spectra used in this study.  The spectra were individually velocity-shifted to remove the RV variation from the orbital reflex motion.  The average signal-to-noise ratio is approximately 270 per resolution element.  }
\label{fig:spectra}
\end{figure*} 
 
 \newpage
\section{The Pulse Timings of CS 1246}
\label{O-C}

Using the 0.41-m PROMPT array, we aggressively monitored the pulsations of CS 1246 for four consecutive months in 2010 to look for secular variations in the pulse timings.  The phase of the 371.7-s pulsation mode (the only confirmed mode) was used to construct an O-C diagram, which revealed a parabolic secular trend with overlying sine wave.  For full details on these observations, including the construction and interpretations of the O-C diagram, we refer the reader to \citet{bar11}.  

A secular decrease in the pulsational period on the order of 1 ms every 1.7 years can explain the parabolic component.  We must note, however, it is plausible that the quadratic term is instead a sine wave with period much longer than our baseline, which could masquerade as a parabola until a wider phase coverage is obtained.

The sinusoidal trend is undoubtedly more convincing.  Having undergone 30 cycles over the course of our observing baseline, the phase oscillation appeared in both the amplitude spectrum of the combined light curve (as a splitting of the main mode) and in a Fourier transform of the O-C diagram itself.  Several phenomena can be invoked to account for this variation, but the presence of a previously-unknown binary companion provided the simplest explanation.  As the orbiting body tugs on CS 1246, the light-travel distance changes along our line-of-sight, resulting in a phase oscillation of the pulsations.  This hypothesis is easily falsifiable since it makes the following prediction:  if an orbital reflex motion is the cause of the phase oscillation, CS 1246 should exhibit line-of-sight radial velocity variations with a semi-amplitude of 16.6 km s$^{-1}$ and period of two weeks.  Moreover, the variation should be 90 degrees out-of-phase with the oscillation in the O-C diagram.  Table 1 presents the full set of orbital parameters predicted from the pulse timings, as published by \citet{bar11}.

\begin{table}
\begin{tabular}{cllll}
\multicolumn{5}{c}{TABLE 1}\\
\multicolumn{5}{c}{\textsc{Orbital Parameters from the O-C Diagram}}\\
\hspace{3pt}\\
\hline
\hline
\textbf{Param} & \textbf{Value} & \textbf{Error} & \textbf{Units} & \textbf{Comments}\\
\hline
$\Pi$ & 14.105 & $\pm$ 0.011 & days & orbital period\\
$K_1$ & 16.6 & $\pm$ 0.6 & km s$^{-1}$ & CS1246 RV semi-amp\\
$f$ & 0.0066 & $\pm$ 0.0007 & M$_{\sun}$ & mass function\\
$a$ & 0.0963 & $\pm$ 0.0003 & AU & separation distance\\
$m  sin i$$^a$ & 0.129 & $\pm$ 0.005 & M$_{\sun}$ & companion mass\\
\hline
\multicolumn{5}{l}{$^a$\begin{scriptsize}assuming the canonical sdB mass 0.47 M$_{\sun}$ for CS 1246\end{scriptsize}}
\end{tabular}
\label{tab:O-C_prediction}
\end{table}

\section{Spectroscopy}
\label{obs}

We began taking spectra of CS 1246 in January 2011 using the Goodman spectrograph on the 4.1-m SOAR telescope.  Appendix A presents the details of our observations.  Since an RV oscillation with an observed semi-amplitude of 8.8 km s$^{-1}$ is generated by the pulsational motions themselves \citep{bar10},  we set the exposure time of each spectrum to the pulsation period, 371.7-s.  This technique washes out the pulsational RV variations and makes it easier to detect changes in the systemic velocity.  We used a 0.46\arcsec\ slit and a 2100 mm$^{-1}$ VPH grating from Syzygy Optics, LLC, to cover the spectral range 3690-4390 \AA\ with a resolution of 0.52 \AA\  (0.17 \AA\ per pixel).  To avoid contaminating the spectra with light from a dim star a few arcseconds to the West, we set the position angle along the slit to 270 degrees East of North and sacrificed observing at the parallactic angle.  Flexure in the spectrograph as an object is tracked changes the wavelength solution slightly over several hours; consequently, we avoided taking more than 30 minutes of stellar spectra in a row without obtaining comparison spectra of an Fe lamp.  When time permitted, we also observed an A-type radial velocity standard (HR 3383; \citealt{fek99}) to determine the velocity zero-point.  In total, we took 149 individual spectra of CS 1246 over a six-month period.

Each spectral frame was bias-subtracted and flat-fielded in IRAF.  Additionally, we subtracted ``pseudo" darks to remove a coherent source of stray light in the spectrograph with a count rate of nearly six photons per minute per pixel.  These calibration frames were obtained by placing the spectrograph in the same configuration used for the target observations and integrating with the shutter open (in darkness, with the flip-mirror of the telescope closed).  We did not flux-calibrate the spectra as time did not permit us to observe spectroscopic standard stars each night.

We combined together all spectra taken consecutively on a given night to produce ``master" spectra.  On most nights, we were able to produce two master spectra, with total integration times around 26 minutes and signal-to-noise ratios of 60 per resolution element, on average.  Figure \ref{fig:spectra} presents an example of one of these spectra plotted above the mean spectrum created using all of our data.  

\section{The Radial Velocity Curve}
\label{RV}

We employed two methods to measure the radial velocities in our master spectra.   Both are iterative processes during which we (i) create a mean spectrum from all individual spectra, (ii) measure velocities of individual spectra based off of information from the mean, (iii) create a new mean spectrum by velocity-shifting the individual spectra, and (iv) repeat the above steps until the solution converges.  Further details on these methods follow.\begin{description}
\item[\textbf{(1) Lorentzian+Gaussian Fits to H Balmer Lines}  ]  Using the MPFIT routine in IDL \cite{mar09}, which employs the Levenberg-Marquardt method, we first fit the sum of a Gaussian and Lorentizian to the individual H Balmer line profiles in the mean spectrum.  Fixing the shapes of the profiles to those fit in the mean spectrum, we then fit the Balmer line profiles in the individual spectra and used the best-fit centroid values to compute velocities.  The mean velocity of each master spectrum was calculated from the weighted-average of the Balmer line velocities.  We created an improved template by de-shifting the individual spectra using the previous RV results and repeated the process until the solution converged.  Balmer lines beyond H9 were ignored since the low signal-to-noise ratio in this part of the spectrum resulted in inconsistent fits.  We estimated the velocity error in each master spectrum using the standard deviation of the individual line-profile measurements.  If a night contained more than one master spectrum, we averaged the resulting measurements.  To check the zero-point, we also measured the radial velocity of the velocity standard HR 3383.  As our measurement agreed with the published value to within the errors, we did not apply any offsets to our measured velocities for CS 1246.
\item[\textbf{(2) Cross-Correlation} ]  Following a procedure similar to \citet{saf01}, we used the \textit{fxcor} routine in IRAF, which uses the technique of \citet{ton79}, to compute the velocity shift of each master spectrum.  All spectra were combined into a template that was cross-correlated against the individual, unshifted spectra.  An improved template was then created by de-shifting the individual spectra using the previous cross-correlation results, and the process was repeated until the solution converged.  We converted the relative velocities to absolute velocities by cross correlating the template with the spectrum of our observed radial velocity standard, HR 3383.  Errors were taken from the output of the \textit{fxcor} task in IRAF.  If a night contained more than one master spectrum, we averaged the resulting RV measurements. 
\end{description}  After determining the velocities, we used the IRAF task \textit{rvcorrect} to calculate the heliocentric corrections at our observing epochs and applied these corrections to our measurements.  Results are shown in Appendix A.

\begin{figure*}

  \centering
  	(a)
  {
     \plotone{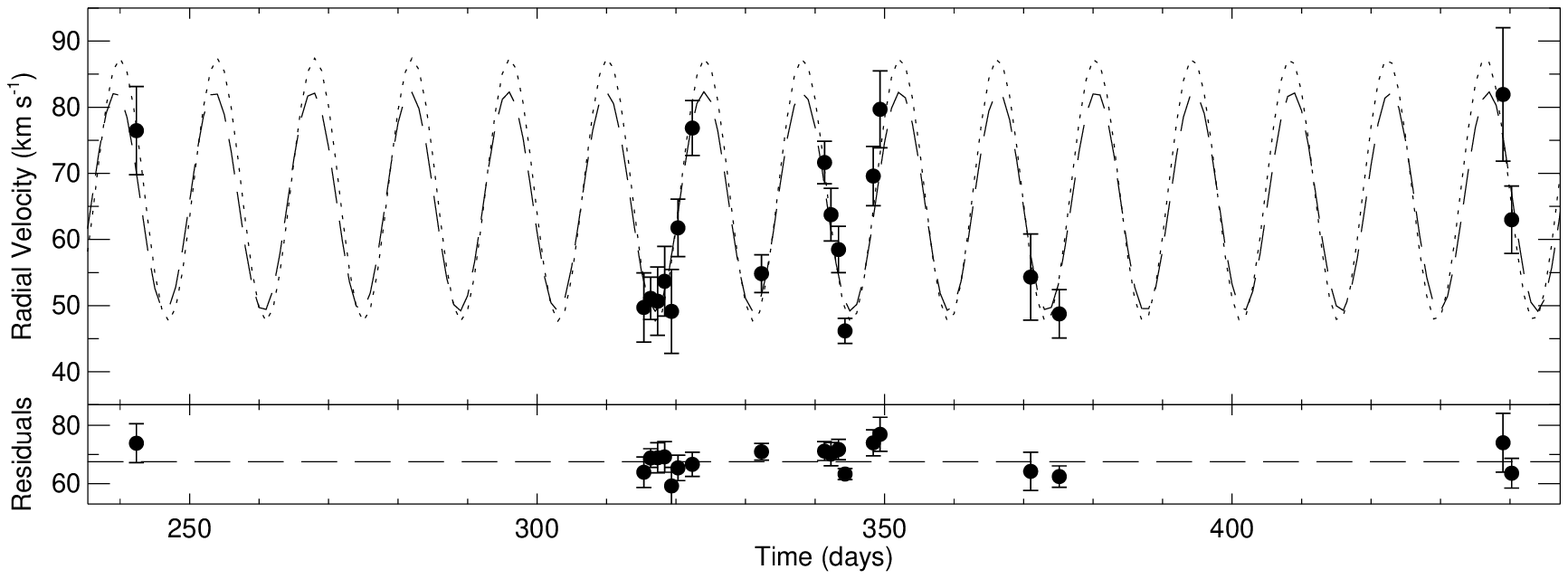}
     
  }
 (b)
  {
     \plotone{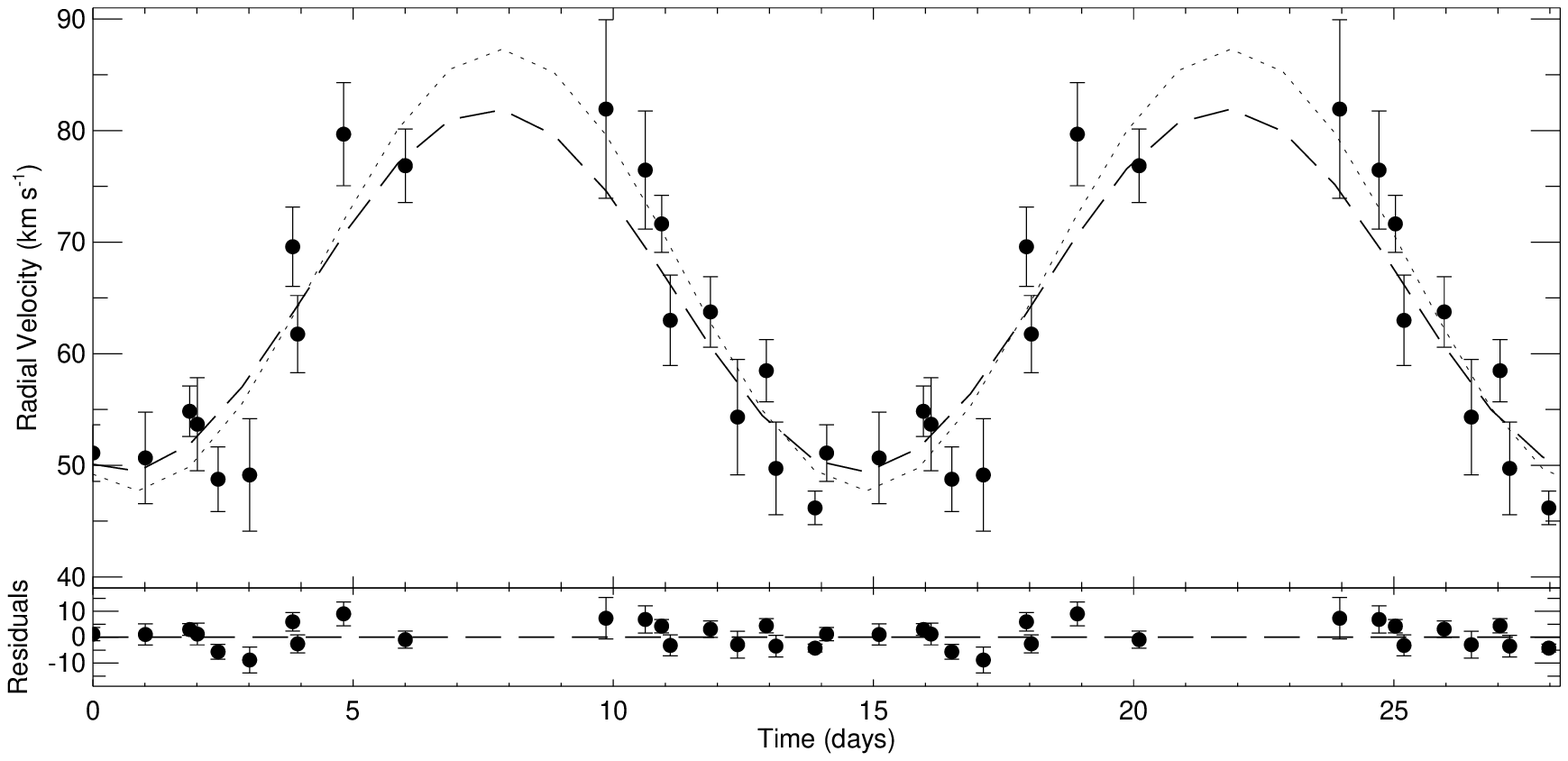}
     
  }
  
  \caption{Heliocentric radial velocities of CS 1246. \textbf{(a)} \textit{Top panel:}  RV measurements derived from Lorentzian+Gaussian fits to the H Balmer absorption-line profiles.  The dashed line marks the velocity curve inferred from the phase oscillation in the O-C diagram, under the assumption it is caused by orbital reflex motion.  Note the agreement in the period, amplitude, and phase.  The dotted line shows the best-fitting sine wave to the data, with all parameters left free.  \textit{Bottom panel:}  Residuals after subtracting from the data the RV curve predicted by the O-C diagram.  \textbf{(b)} \textit{Top panel:} RV curve folded on the period predicted by the O-C diagram (14.105 days), plotted twice for better visualization.  The dashed and dotted lines show the O-C diagram prediction and the best-fit sine wave, respectively.  \textit{Bottom panel:} Residuals after subtracting from the data the RV curve predicted by the O-C diagram.  The standard deviation is 4.7 km s$^{-1}$} 
  \label{fig:raw_rv}
\end{figure*}

RV curves produced using both methods display similar variations.  Figure \ref{fig:raw_rv}a presents the curve from the Lorentzian+Gaussian method as an example.  The dashed line marks the variation predicted by the O-C diagram, with the overall systemic velocity fit as a free parameter.  Figure \ref{fig:raw_rv}b shows the same measurements phase-folded on the 14.1-day periodicity observed in the O-C diagram.

Although inspection of these figures makes clear the agreement between data and predictions, we performed non-linear, least-squares fits of sine waves to the RV curves to quantify the similarities, leaving all parameters free.  Table 2 presents the best-fit parameters above the O-C diagram predictions; the dotted lines in Figure \ref{fig:raw_rv} show the best-fitting sine wave to the Lorentzian+Gaussian fit RV points.   The period and amplitude calculated using both methods agree with those predicted by the O-C diagram.  Our RV data allow us to calculate for the first time the system velocity.  By fitting a sine wave to the data with the period, amplitude, and phase fixed to their O-C calculated values, we derive a systemic velocity of 67.3 $\pm$ 1.7 km s$^{-1}$.

\section{Conclusions}
\label{analysis}

We have presented radial velocity data confirming the binary nature of CS 1246.  The period, amplitude, and phase of the RV variation agree with those predicted by the O-C diagram when interpreting the observed phase oscillation as an orbital reflex motion effect.  Although this is not the first instance such an agreement between pulse timing and velocity results has been noted, this study is the first using pulsations on the order of minutes and the first for a hot subdwarf star.  Our result demonstrates that the rapid pulsations of sdB stars can be adequate clocks for detecting binary companions using the light-travel time technique.  

Since the error bars of the orbital parameters calculated from the O-C diagram fits are smaller than those calculated from the RV curve, our characterization of the companion remains the same as originally presented in \citet{bar11} (see Table 1).  Using the canonical sdB mass (0.47 M$_{\sun}$) for CS 1246, we calculated the minimum mass of the companion to be 0.13 M$_{\sun}$.  Since no eclipses or other photometric indications of the companion have been observed, the orbital inclination angle remains unknown, and we are limited to computing its minimum mass.  There is a 95\% probability the companion mass is less than 0.45 M$_{\sun}$, assuming an isotropic distribution of inclination angles.  Thus, CS 1246 is most likely orbited by a low-mass white dwarf or M-type main sequence star.  

According to the binary population synthesis models of \citet{han02,han03}, an sdB binary with a two-week orbital period could be the product of both common envelope ejection and stable Roche lobe overflow formation channels.  The latter is unlikely for the CS 1246 system since it predicts an early-type main sequence companion, signatures of which would be visible in our spectra but are not.  Both the ``first" and ``second" common envelope ejection channels presented in \citet{han02,han03} are more plausible.  In these scenarios, the sdB progenitor, a red giant star, transfers mass dynamically to its companion prior to reaching the tip of the giant branch.  Eventually a common envelope forms, the orbit shrinks due to friction and tidal interactions, and the envelope of the giant star is ejected, leaving behind a close binary composed of an sdB star with a main sequence (``first" channel) or white dwarf (``second" channel) companion.  Their models predict similar sdB masses for both companion types.  If the companion is an M-dwarf, the system might show an infrared excess, but looking for this photometric signature is complicated by the presence of the Coalsack Dark Nebula, which reddens the light significantly.  Due to these complications, we are left with no choice but to continue improving measurements of the orbital parameters of this system using additional pulse timing data.
 
 \begin{table}
\begin{tabular}{rcccc}
\multicolumn{4}{c}{TABLE 2}\\
\multicolumn{4}{c}{\textsc{Velocity Curve Fit Parameters}}\\
\hspace{3pt}\\
\hline
\hline
\textbf{Method} & \textbf{Period} & \textbf{Amplitude} & \textbf{Systemic RV}\\
\ & (days) & (km s$^{-1}$) & (km s$^{-1}$)\\
\hline
Lorentz+Gauss & 14.03 $\pm$ 0.06 & 19.8 $\pm$ 2.7 & 67.5 $\pm$ 1.9\\ 
Cross-Corr. & 13.92 $\pm$ 0.15 & 13.9 $\pm$ 2.9 & 66.9 $\pm$ 2.7\\
\hline
O-C Diagram & 14.104 $\pm$ 0.011 & 16.6 $\pm$ 0.6 & \textit{no prediction}\\
\hline
\end{tabular}
\label{tab:rv_fit}
\end{table}
 
\begin{acknowledgements}
We acknowledge the support of the National Science Foundation, under award AST-0707381, and are grateful to the Abraham Goodman family for providing the financial support that made the spectrograph possible.  We also recognize the observational support provided by the SOAR operators Alberto Pasten, Patricio Ugarte, Daniel Maturana, and, most notably, Sergio Pizarro, who kindly observed for us during a network outage.  BNB would personally like to thank Steve Heathcote for Director's Discretionary Time and Sheila Kannappan and her group for graciously providing some of their twilight time for this project.  BNB also acknowledges the Rivendell Cabin in the Uwharrie Mountains of North Carolina for providing a peaceful location in which to formulate the layout of this manuscript.  The SOAR Telescope is operated by the Association of Universities for Research in Astronomy, Inc., under a cooperative agreement between the CNPq, Brazil, the National Observatory for Optical Astronomy (NOAO), the University of North Carolina, and Michigan State University, USA.
\end{acknowledgements}



\begin{appendix}

\begin{table}[h]
\centering
\begin{tabular}{cccc}
\multicolumn{4}{c}{TABLE A1}\\
\multicolumn{4}{c}{\textsc{Log of Master Spectra \& Velocities}}\\
\hline
\hline
\textbf{Date}$^a$  & \textbf{HJD}$^a$ &\textbf{Integration Time} & \textbf{Radial Velocity}$^b$\\
 (UT) & (HJD) & (s)  & (km s$^{-1}$)\\
\hline
2010-11-09 &2455509.85674  &4 $\times$ 371.7  & 76.5  $\pm$ 4.9\\
2011-01-21  &2455582.85745 & 7 $\times$ 371.7   &  49.7 $\pm$  4.2\\
2011-01-22   &2455583.83098   &8 $\times$ 371.7   & 51.1 $\pm$ 2.6\\
2011-01-23   & 2455584.83992 & 10 $\times$ 371.7   & 50.7 $\pm$ 4.0\\ 
2011-01-24    &2455585.83775 & 10 $\times$ 371.7   & 53.7 $\pm$ 4.2\\
2011-01-25   &2455586.84322   & 9 $\times$ 371.7   & 49.1  $\pm$ 5.0\\
2011-01-26    &2455587.76537 & 16 $\times$ 371.7   & 61.8 $\pm$ 3.4\\
2011-01-28  &2455589.83534 & 11 $\times$ 371.7   & 76.9  $\pm$ 3.2\\
2011-02-07   & 2455599.78876   & 6 $\times$ 371.7  & 54.8   $\pm$ 2.2\\
2011-02-16  & 2455608.85764 & 6 $\times$ 371.7  &  71.6 $\pm$ 2.6\\
2011-02-17  & 2455609.79579  & 8 $\times$ 371.7  & 63.8 $\pm$ 3.2\\
2011-02-18    & 2455610.86934   &8 $\times$ 371.7   &58.5 $\pm$ 2.7\\
2011-02-19   & 2455611.80540   & 12 $\times$ 371.7  & 46.2 $\pm$ 1.4\\
2011-02-23   & 2455615.86702  &  5 $\times$ 371.7  & 69.6  $\pm$ 3.5\\
2011-02-24   & 2455616.84545  & 6 $\times$ 371.7  & 79.7  $\pm$ 4.6\\
2011-03-18   & 2455638.51437   &10 $\times$ 371.7  &  55.3 $\pm$ 5.1\\
2011-03-22   & 2455642.63103 & 7 $\times$ 371.7  &  48.8 $\pm$ 2.9\\
2011-05-24  & 2455706.48698  & 3 $\times$ 371.7 & 81.9  $\pm$ 7.8\\
2011-05-26  & 2455707.72320 & 3 $\times$ 371.7 & 63.0  $\pm$ 4.0 \\
\hline
\multicolumn{4}{l}{$^a$ at mid-integration}\\
\multicolumn{4}{l}{$^b$ heliocentric velocities from Lorentzian+Gaussian fitting method}\\
\end{tabular}
\label{tab:obs}
\end{table}

\end{appendix}




\end{document}